 \documentclass[smallabstract,smallcaptions]{dccpaper}

\usepackage{epsfig}
\usepackage{citesort}
\usepackage{amsmath}
\usepackage{amssymb}
\usepackage{color}
\usepackage{url,xspace,graphicx}
\usepackage{caption}
\usepackage{subcaption}

\newlength{\figurewidth}
\newlength{\smallfigurewidth}

\newcommand{\our}{\textsc{GCIS}\xspace}
\newcommand{\SAIS}{\textsc{SAIS}\xspace}

\newcommand{\repair}{\textsc{Re-Pair}\xspace}
\newcommand{\gzip}{\textsc{GZIP}\xspace}

\newcommand{\szip}{\textsc{7-zip}\xspace}

\newcommand{\lcp}{\texttt{lcp}\xspace}
\newcommand{\conc}{\cdot}
\newcommand{\etal}{{\it et al.}\xspace}
\newcommand{\F}{\mathcal{F}}
\newcommand{\suff}{\mathrm{s}}

\newtheorem{definition}{Definition}

\newcommand{\SA}{\ensuremath{\mathsf{SA}}\xspace}

\begin{document}

\title
{\large
\textbf{A Grammar Compression Algorithm based on Induced Suffix~Sorting}
}

\author{
Daniel Saad Nogueira Nunes$^{1,2}$, Felipe A. Louza$^{3~\ast}$, \\ 
Simon Gog$^{4}$, Mauricio Ayala-Rinc\'on$^{2}$ and Gonzalo Navarro$^{5}$ \\[1.0em]
\hspace*{-0.2in}
{\small\begin{minipage}{\linewidth}\begin{center}
$^{1}$Federal Institute of Education, Science and Technology of Brasília
\url{daniel.nunes@ifb.edu.br}\\
$^{2}$Department of Computer Science, Brasilia University, Brazil \\
\url{ayala@unb.br} \\
$^{3}$Department of Computing and Mathematics, University of S\~ao Paulo, Brazil \\
\url{louza@usp.br}\\
$^{4}$Institute of Theoretical Informatics, Karlsruhe Institute of Technology, Germany \\
\url{gog@kit.edu}\\
$^{5}$Department of Computer Science, University of Chile, Chile\\
\url{gnavarro@dcc.uchile.cl} \\
\end{center}\end{minipage}}
\\[1.0em]
\thanks{$^{\ast}$
FAL was supported by the grant $\#$2017/09105-0 from the S\~ao Paulo Research Foundation (FAPESP).
}
}


\maketitle
\thispagestyle{plain}
\pagestyle{plain}

\begin{abstract} 
We introduce \our, a grammar compression algorithm based on 
the induced suffix sorting algorithm \SAIS, introduced by Nong \etal in 2009.
Our solution builds on the factorization performed by \SAIS during suffix sorting.
We construct a context-free grammar on the input string which can be further reduced into a
shorter string by substituting each substring by its correspondent factor.
The resulting grammar is encoded by exploring some redundancies, such as common prefixes between
suffix rules, which are sorted according to \SAIS framework. 
When compared to  well-known compression tools such as \repair and \szip, our algorithm is competitive and very effective at handling repetitive string regarding compression ratio, compression and decompression running time. 

\end{abstract}

\Section{Introduction}

Text compression consists in transforming an input string into another string
whose bit sequence representation is smaller.
Given the suffix array~\cite{Manber1993,Gonnet1992} of a string, one
can compute efficiently the Burrows-Wheeler transform
(BWT)~\cite{Burrows1994} and the Lempel-Ziv factorization
(LZ77)~\cite{Ziv1977,Ohlebusch2011,Karkkainen2013,Goto2014}, which are at the heart of
the popular data compression tools \szip and \gzip~\cite{Navarro2016}.


In 2009, Nong \etal~\cite{Nong2009a} introduced a remarkable algorithm
called \SAIS, which runs in linear time and is fast in practice
to construct the suffix array. 
Subsequently, \SAIS was adapted to compute directly the
BWT~\cite{Okanohara2009}, the $\Phi$-array~\cite{Karkkainen2009,Goto2014}, the
LCP array~\cite{Fischer2011}, and the suffix array for string
collections~\cite{Louza2017c}.


In this article we introduce \our, a new grammar-based compression algorithm
that builds on \SAIS.
We construct a context-free grammar based on the 
string factorization performed by \SAIS recursively.
The rules are encoded according to the length of longest common prefixes between
consecutive rules, which are sorted lexicographically by \SAIS.

Experiments have shown that, regarding repetitive strings, \our is competitive with 
\repair~\cite{Larsson1999b} and \szip~\cite{pavlov}, since \our presents the fastest
compression time, while maintaining a compression ratio close to \repair, but
being the slower to decode. Hence, it is a practical alternative when considering all trade-off aspects. Moreover GCIS utilizes a novel grammar compression framework in the sense that it is the first, as far as the authors are concerned, based on induced suffix sorting.


\Section{Background}\label{s:background}

Let $T$ be a string of length $|T|=n$, $T = T[1,n] = T[1]\conc T[2]\dots \conc T[n]$, over
a fixed ordered alphabet $\Sigma$.
A constant alphabet has size $\sigma \in O(1)$ and an integer alphabet has size $\sigma \in n^{O(1)}$.
We denote the concatenation of strings or symbols by the dot operator
($\conc$), which can be omitted.
We use the symbol $<$ for the lexicographic order relation between strings.

For convenience, we assume that $T$ always ends with a special symbol
$T[n]=\$$, which is not present elsewhere in $T$ and lexicographically precedes every symbol
in $\Sigma$.
Let $T[1,j]$ be the prefix of $T$ that ends at position $j$, and $T[i,n]$ be the suffix 
of $T$ that starts at position $i$ also denoted as $T_i$ by brevity.
We denote the length of the longest common prefix of two strings $T_1$ and $T_2$ in $\Sigma^*$
by $\lcp(T_1,T_2)$.

The suffix array (\SA)~\cite{Manber1993,Gonnet1992} of a string $T[1,n]$ is an array
of integers in the range $[1, n]$ that gives the lexicographic order of all
suffixes of $T$, such that $T_{\SA[1]} < T_{\SA[2]} < \ldots < T_{\SA[n]}$.
The suffixes starting with the same symbol $c \in \Sigma$ form a $c$-bucket in the
suffix array.
The head and the tail of a bucket refer to the first and the last position of
the bucket in \SA.

Let $G = (\Sigma, \Gamma, P, X_S)$ be a reduced context-free grammar (does not contain unreachable non-terminals). $\Sigma$ is the terminal alphabet of $G$;
$\Gamma$ is the set of non-terminals symbols that is disjoint from $\Sigma$; 
$P \subseteq \Gamma \times (\Sigma \cup \Gamma)^*$ is the set of production rules; and
$X_S \in \Gamma$ is the start symbol.
A production rule $(X_i, \alpha_i)$ is also denoted by $X_i \rightarrow \alpha_i$.
We say that $\alpha_i$ is derived from $X_i$.
For strings $s,t \in (\Sigma \cup \Gamma)^*$, we say that $t$ derives from $s$ if it is obtained by application of a production rule in $P$; we say that $t$ is generated from $s$ if $t$ is obtained by a sequence of derivations from $s$. We define $|G|$ as the total length of the strings on the right side of all rules.

Given a string $T$, grammar compression is to find a grammar $G$ which generates only $T$ such that $G$ can be represented in less space than the original $T$.
Given that $G$ grammar-compresses $T$, for $(X_i,\alpha_i)\in P$, we define $\F(X_i)=s$ as the single string $s\in\Sigma^*$ that is generated from $\alpha_i$. 
The language generated by $G$ is $L(G) = \F(X_S)$.




\Section{Related work}\label{s:related} 

\SAIS~\cite{Nong2009a} builds on the induced suffix sorting technique
introduced by previous algorithms~\cite{Itoh1999,Ko2003}.
Induced suffix sorting consists in deducing the order of unsorted suffixes from
a set of already ordered suffixes.

The next definition classifies suffixes and symbols of  strings.

\begin{definition}[L-type and S-type]
For any string $T$, $T_n=\$$ has type S.
A suffix $T_i$ is an S-suffix if $T_i < T_{i+1}$, otherwise $T_i$ is an L-suffix. $T[i]$ has the type of $T_i$. 
\end{definition}

The suffixes can be classified in linear time by scanning $T$ once from
right to left.
The type of each suffix is stored in a bitmap of size $n$. 

Note that, in a $c$-bucket, all L-suffixes precede to the S-suffixes.

Further, the classification of suffixes is refined as below: 

\begin{definition}[LMS-type]
Let $T$ be a string. $T_i$ is an LMS-suffix if $T_i$ is an S-suffix and $T_{i-1}$ is an L-suffix. 
\end{definition}

Nong \etal~\cite{Nong2009a} showed that the order of the LMS-suffixes is enough
to induce the order of all suffixes.

\SAIS works as follows:

\SubSection{SAIS framework:}

\begin{enumerate}

\item Sort the LMS-suffixes. This step is explained below.

\item Insert the LMS-suffixes into their $c$-buckets in \SA, without changing
their order.

\item Induce L-suffixes by scanning \SA from left to right: for each suffix
$\SA[i]$ if $T[\SA[i]-1]$ is L-type, insert $\SA[i]-1$ into the 
head of its bucket. 

\item Induce S-suffixes by scanning \SA from right to left: for each suffix
$\SA[i]$ if $T[\SA[i]-1]$ is S-type, insert $\SA[i]-1$ into the
tail of its bucket. 

\end{enumerate}

We say that whenever a value is inserted in the head (or tail) of a bucket,the head (or tail) is increased (or decreased) by one.



In order to sort the LMS-suffixes in Step 1, $T[1,n]$ is divided (factorized)
into LMS-substrings.

\begin{definition}\label{def:lms-type}
$T[i,j]$ is an LMS-substring if both $T_i$ and $T_j$ are LMS-suffixes,
but no suffix between $i$ and $j$ has LMS-type.
The last suffix $T_n$ is an LMS-substring.
\end{definition}

Let $r^1_1, r^1_2, \dots, r^1_{n^1}$ be the $n^1$ LMS-substrings of $T$ read from left-to-right.
A modified version of \SAIS is applied to sort the LMS-substrings.
Starting from Step 2, $T[1,n]$ is scanned (right-to-left) and each unsorted LMS-suffix is
inserted (bucket-sorted) regarding its first symbol at the tail of its $c$-bucket. 
Steps 3 and 4 work exactly the same.
At the end, all LMS-substrings are sorted and stored in their corresponding
c-buckets in \SA.

\SubSection{Naming:}

A {\it name} $v^1_i$ is assigned to each LMS-substring $r^1_i$ according to its
lexicographical rank in $[1, \sigma^1]$, such that $v^1_i < v^1_j$ if $r^1_i < r^1_j$,
$v^1_i = v^1_j$ if $r^1_i = r^1_j$ and $\sigma^1$ is the number of different
LMS-substrings in $T$.
In order to compute the names, 
each consecutive LMS-substrings in \SA, say $r^1_i$ and $r^1_{i+1}$, are compared 
to determine if either $r^1_i = r^1_{i+1}$ or $r^1_i < r^1_{i+1}$.
In the former case $v^1_{i+1}$ is named as $v^1_i$, whereas in the latter case $v^1_{i+1}$ is named as $v^1_i+1$. 
This procedure may be sped up by comparing the LMS-substrings first by symbol and then by
type, with L-type symbols being smaller than S-type symbols in case of
ties~\cite{Nong2011}.

\SubSection{Recursive call:}

A new (reduced) string $T^1 = v^1_1 \conc v^1_1 \cdots v^1_{n^1}$ is created, whose
length ${n^1}$ is at most $n/2$, and the alphabet size $\sigma^1$ is integer.
If every $v^1_i\ne v^1_j$ then all LMS-suffixes are already sorted.
Otherwise, \SAIS is recursively applied to sort all the suffixes of $T^1$.
Nong \etal~\cite{Nong2009a} showed that the relative order of the LMS-suffixes
in $T$ is the same as the order of the respective suffixes in $T^1$.
Therefore, the order of all LMS-suffixes can be determined by the result of the
recursive algorithm.


%

\Section{Grammar Compression by Induced Suffix Sorting}\label{s:algorithm}

In this section we introduce the grammar compression by induced sorting
(\our), which is based on \SAIS.

First, we compute a context-free grammar $G = (\Sigma, \Gamma, P, X_S)$ that
generates only $T[1,n]$.  
To do this we modify \SAIS as follows.

\SubSection{Grammar construction:}


Considering the $j$-th recursion level, in Step 1, after the input string $T^j[1,n]$ is divided into the LMS-substrings
$r^j_1, r^j_2, \dots, r^j_{n^j}$ and named into
$v^j_1, v^j_2, \dots, v^j_{n^j}$, 
we create a new rule $X_i \rightarrow \alpha_i$ for each
different LMS-substring $r^j_i = T^j[a,b]$ in the form
$r(v^j_i) \rightarrow T^j[a,b-1]$, where $r(v^j_i) = v^j_i + \sum_{k=1}^{j-1} \sigma^k$.
Moreover, we create an additional rule $r(0^j) \rightarrow T^j[1,j_1-1]$ for the
prefix of $T^j$ that is not included in the first LMS-substring $r^j_1$.

The algorithm is called recursively with the reduced string $T^{j+1} = v^j_1 \conc
v^j_2 \cdots v^j_{n^j}$ as input while $\sigma^j<n^j$, that is, the LMS-substrings are not pairwise distinct.
At the end, when $\sigma^j=n^j$, we create the start symbol of $G$ as being
$X_S$, such the production $X_S \rightarrow r(0^j) \conc r(v^j_1) \conc r(v^j_2) \cdots r(v^j_{n^j})$
generates only the original string $T[1,n]$.

The algorithm stops after computing $X_S$, since we are not interested in constructing
the suffix array,  we do not execute Steps 2, 3 and 4 of
\SAIS. 
The recursive calls return to the top level and we have computed a grammar $G$
that generates only $T[1,n]$.

Since for each LMS-substring a unique $r(v^j_i)$ exists, there are no cycles in any derivations, and $L(G)=T$, we have that $G$ is a grammar that compresses $T$ \cite{arpeR06}.

\SubSection{Grammar compression:}

Consecutive entries in the set of productions $P$ are likely to share a common
prefix, since the LMS-substrings are given lexicographically ordered by \SAIS.
Therefore, each rule $X_i \rightarrow \alpha_i \in P$ is encoded using two
values $(\ell_i, \suff(\alpha_i))$, such that $\ell_i$ encodes the length of
longest common prefix (\lcp) between $\alpha_{i-1}$ and $\alpha_{i}$, and the
remaining symbols of $\alpha_{i}$ are given by
$\suff(\alpha_i)=\alpha_i[\ell_i+1,|\alpha_i|]$.
This technique is known as Front-coding~\cite{Witten1999}.

The computation of $(\ell_i, \suff(\alpha_i))$ is performed with no additional
cost with a slight modification in the naming procedure of \SAIS.
Each consecutive LMS-substring in \SA, say $r^j_{i-1}$ and $r^j_{i}$ are compared first
by symbol and then by type to check if either $r^j_{i-1} = r^j_{i}$ or $r^j_{i-1} < r_{i}$.
In order to compute $\lcp(r^j_{i-1},r^j_{i})$ we compare them only by symbol until
finding the first mismatch.
The resulting order is the same with a small slowdown in the running time.


\SubSection{Computational cost:}

\our runs in $O(n)$ time, since each step of the modified \SAIS is linear and
the length of the reduced string $T^j$ is at most $|T^{j-1}|/2$.

\Section{Implementation details}\label{s:implementation}

In this section we discuss implementation details of the \our encoding and decoding
processes.

\SubSection{Encoding:}

A rule $X_i$ is derived into a pair $\alpha_i = (\ell_i,\suff(\alpha_i))$, where $\ell$ equals $\lcp(\alpha_{i-1},\alpha_{i})$  and $\suff(\alpha_{i})$ corresponds to the remaining $\alpha_i[\ell_i+1,|\alpha_{i}|]$ symbols. The $\ell$ values tend to be small and, considering the $j$-th recursion value, the sum of such values cannot be greater than $n^j$, since no two LMS-substrings overlap.

One can encode all $\ell$ values into a sequence of computer words $L$ by using {\tt Simple8b} encoding \cite{AnhM10}. This technique packs a number of small integers in a $64$-bit word using the number of bits required by the largest integer. Basically it identifies a word with a $4$-bit tag called \textit{selector}, which specifies the number of integers encoded in a single word and the width of such integers. {\tt Simple8b} also has specific selectors for a run consisting of zeroes. If a run of $240$ or $120$ zeros is encountered, it can be represented with a single $64$ bit word. Table \ref{tab:s8b} contain all possible selector values, which reflects the possible arrangements of fixed-width integers storage in a single $64$-bit word under this encoding scheme.

\begin{table}[tp]
	\centering
	\caption{{\tt Simple8b} possible arrangements \cite{AnhM10}.}
	\label{tab:s8b}
	\resizebox{\textwidth}{!}{
		\begin{tabular}{lllllllllllllllll}
		\hline
		Selector value & 0   & 1   & 2  & 3  & 4  & 5  & 6  & 7  & 8 & 9 & 10 & 11 & 12 & 13 & 14 & 15 \\ \hline
		Item width     & 0   & 0   & 1  & 2  & 3  & 4  & 5  & 6  & 7 & 8 & 10 & 12 & 15 & 20 & 30 & 60 \\
		Group Size     & 240 & 120 & 60 & 30 & 20 & 15 & 12 & 10 & 8 & 7 & 6  & 5  & 4  & 3  & 2  & 1  \\
		Wasted bits    & 60  & 60  & 0  & 0  & 0  & 0  & 0  & 0  & 4 & 4 & 0  & 0  & 0  & 0  & 0  & 0  \\ \hline
		\end{tabular}
	}
\end{table}

All $\suff(\alpha_{i})$ are encoded in a single fixed-width integer array $R$, consisting of width $\lfloor\lg(\alpha^j)\rfloor+1$ bits. The length of each $\suff(\alpha_{i})$ is also encoded using {\tt Simple8b} into a word array $S$. The same observation of the $\lcp$ sum can be done here: the sum of all $|\suff(\alpha_{i})|$ is no larger than $n^j$.

A greedy strategy was employed to stop the recursion when the dictionary size of the $(j+1)$-th level plus the size in bits of $T^{j+1}$ is bigger than the size in bits of $T^{j}$. In this situation, the computation done on the $j$-th level is discarded and the algorithm stops. When this condition is met, $\alpha^j<n^j$, but this does not interfere on the decoding algorithm. 


\SubSection{Decoding:}

The decoding process is done level-wise, starting from the last level, by decoding the right side of each rule. In the $j$-th level, the values $(x,y,z)$ from $L$, $R$ and $S$ are decoded in a sequential way. In order to compute $\alpha_{k+1}$ from $\alpha_k$, the first $x$ symbols of $\alpha_k$ are copied to $\alpha_{k+1}$ and the $z$ symbols from $R$, which correspond to the string $y$, are copied to $\alpha_{k+1}$ as well.
A bitmap $D$ is built to contain the length of all $\alpha_i$ by using Rice-coding. With two $\textsc{select}_1$ operations it is possible to query the starting point of each $\alpha_i$ in this array and the length $|\alpha_{i}|$ in constant time using $2n^j+o(n^j)$ bits, where $j$ corresponds to the $j$-th recursive step of the grammar construction.

Once all rules are expanded into a fixed-width integer array of $\lfloor \lg(\sigma^j)\rfloor+1$ bits, $T^{j-1}$ can be decoded from $T^j$.
First, the right side of $r(0^j)$ is copied into $T^{j-1}$. Then, $T^j$ is scanned in a left-to-right fashion and for each $T^j[i]$ the algorithm copies  a substring to $T^{j-1}$ which equals the right side of $r(T^j[i])$ and can be easily found with the bitmap $D$ support.

\Section{Experiments}\label{s:experiments}

We compared \our with
\repair\footnote{\url{https://github.com/rwanwork/Re-Pair}} and \szip\footnote{\url{http://p7zip.sourceforge.net/}} regarding
Pizza\&Chili Repetitive
Corpus\footnote{\url{http://pizzachili.dcc.uchile.cl/repcorpus.html}} under the
subjects of compression ratio, compression and decompression running time. 
In particular, we used a space-efficient implementation of \repair by
Wan~\cite{Wan2003}, wich encodes each rule with one integer plus few bits. 
\our was implemented in {\tt C++11} using the Succinct Data Structure Library
(SDSL) \cite{gbmp2014sea}.

All experiments were conducted on machine with {\tt 2x Intel(R) Xeon(R) CPU
E5-2407 v2 @ 2.40GHz} CPUs and $256$GB of RAM memory. The operating system used
was based on the Debian GNU/Linux O.S.
The input size of each experiment is given in the second column of
Tables~\ref{tab:compression-ratio}, \ref{tab:compression-time} and
\ref{tab:decompression-time}.


Experimental results show that our algorithm is very effective
at handling repetitive strings. 
\our presents a competitive compression ratio, compression and decompression
time, being a real practical option when considering all those subjects
simultaneously.

\SubSection{Compression and decompression:} 

Table \ref{tab:compression-ratio} comprises the compression Ratio (\%), 
corresponding to the size of the compressed text over the original input size.
\szip presents the best compression ratio, except for \texttt{coreutils},
\texttt{fib41}, \texttt{rs} and \texttt{tm29}, where \repair outperforms it.
Note that \our presents a competitive compression ratio compared to \repair.

Table~\ref{tab:compression-time} shows the compression time of each algorithm.  
\our is the fastest algorithm, except for \texttt{einstein.de},
\texttt{einstein.en} and \texttt{proteins}, where \szip was the fastest.
\our outperforms \repair and \szip by a large margin in most cases, being up to
$6.5$ times faster than \repair (\texttt{tm29}) and up to $6.9$ times faster
than \szip (\texttt{cere}).


Table~\ref{tab:decompression-time} presents the decompression time of each algorithm.
\szip outperforms \repair and \our, except for \texttt{fib41}, \texttt{rs} and \texttt{tm29}, where \repair was the fastest.
\our is up to $20$ times slower than \repair and \szip (\texttt{einstein.en}), whereas \repair is up to $6.6$ times slower than \szip (\texttt{cere}).

\SubSection{Peak memory}

We evaluated the peak memory consumption of \repair and
\our in compression and decompression procedures. \szip and
was not evaluated since it require negligible amount of space when
compressing or decompressing.  

\begin{figure}[t]
	\centering
	\begin{subfigure}[t]{.475\textwidth}
		\centering
		\includegraphics[scale=.475]{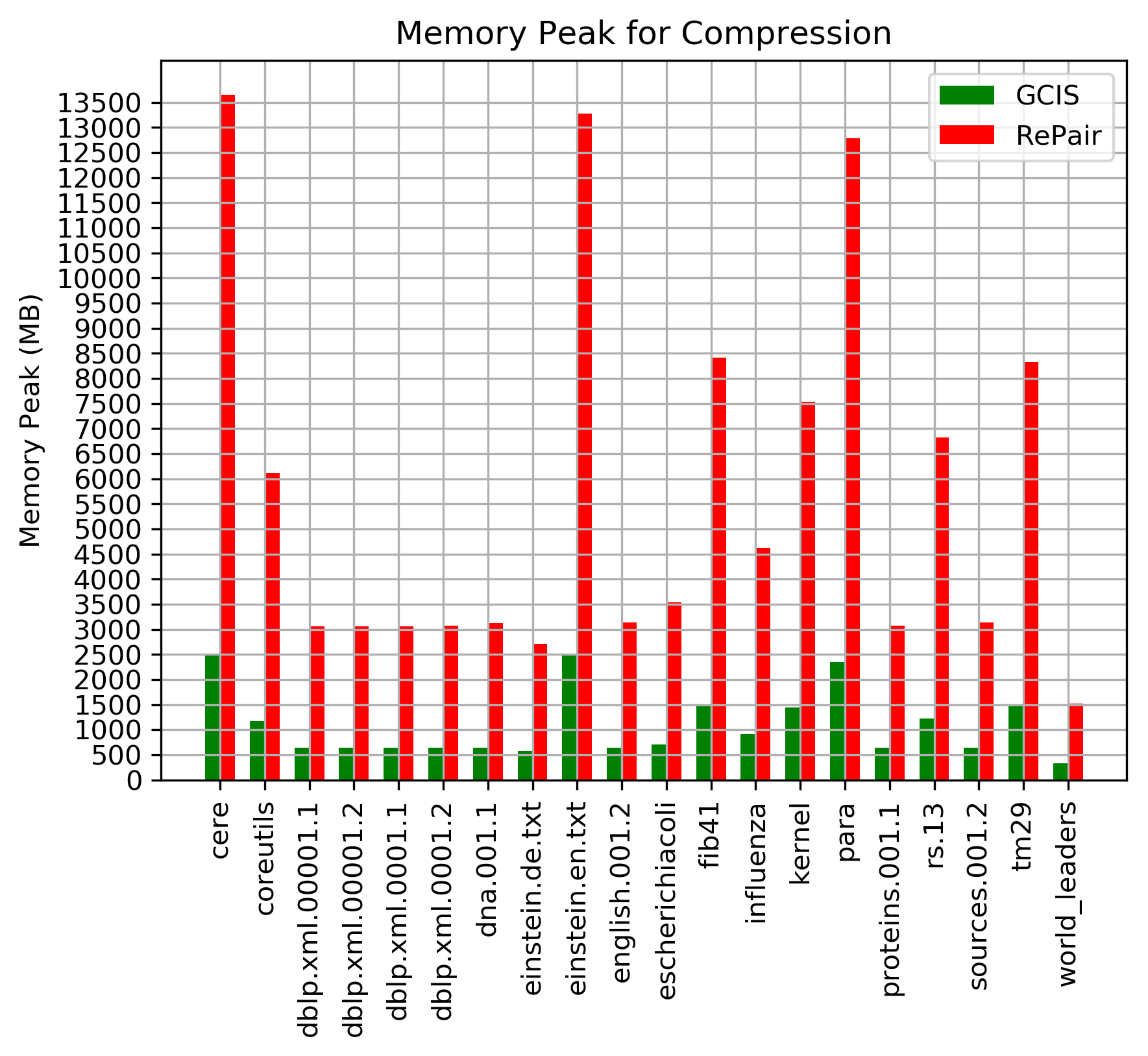}
		\caption{Memory Peak (MB) during compression.}
		\label{fig:pkmem-compress}
	\end{subfigure}
	\begin{subfigure}[t]{.475\textwidth}
		\centering
		\includegraphics[scale=.475]{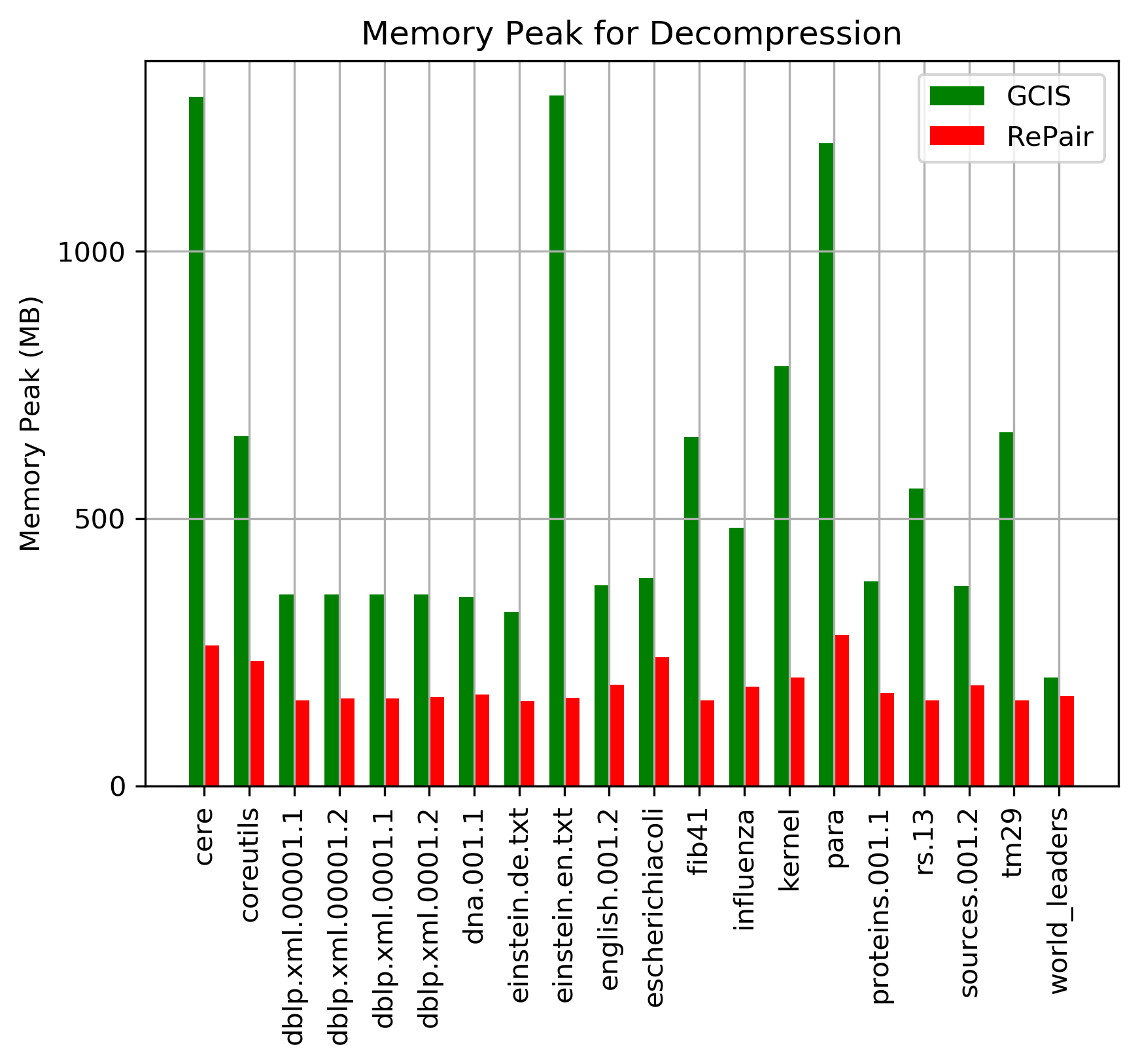}
		\caption{Memory Peak (MB) during decompression.}
		\label{fig:pkmem-decompress}
	\end{subfigure}
	\label{fig:pkmem}
	\caption{Peak memory of \our and \repair regarding compression and decompression.}
\end{figure}

Figure \ref{fig:pkmem-compress} shows that \our
requires five times less the space needed by \repair during compression. Since \our is
based on SAIS, it requires $\approx 5\times n$ bytes, for inputs with $n<4$GB,
whereas \repair requires $\approx 30\times n$ bytes, becoming
prohibitive when the input is large. In decompression, illustrated by
\ref{fig:pkmem-decompress}, \repair has a lower peak memory usage than \our,
making the former more appealing when memory is limited.

\begin{table}[tp]
\centering
\caption{Compression ratio regarding Pizzaz\&Chili repetitive corpus.}
\label{tab:compression-ratio}
\begin{tabular}{|l|c|c|c|c|}
\hline
\multicolumn{2}{|c|}{Pizza\&Chili Repetitive Corpus} & \multicolumn{3}{c|}{Compression Ratio (\%)} \\ \hline
Experiment                & Input Size (MB)          & \our      & \repair    & \szip       	    \\ \hline
\texttt{cere}             & 461.29                   & 3.76      & 1.86      & \textbf{1.82}     \\ \hline
\texttt{coreutils}        & 205.28                   & 5.39      & \textbf{2.54}      & 11.63    \\ \hline
\texttt{dblp.xml.00001.1} & 104.86                   & 0.43      & 0.19     & \textbf{0.16}     \\ \hline
\texttt{dblp.xml.00001.2} & 104.86                   & 0.43      & 0.18      & \textbf{0.16}     \\ \hline
\texttt{dblp.xml.0001.1}  & 104.86                   & 0.84      & 0.46      & \textbf{0.20}     \\ \hline
\texttt{dblp.xml.0001.2}  & 104.86                   & 0.77      & 0.39      & \textbf{0.19}     \\ \hline
\texttt{dna.001.1}        & 104.86                   & 3.55      & 2.43      & \textbf{0.51}     \\ \hline
\texttt{einstein.de.txt}  & 92.76                    & 0.31      & 0.16      & \textbf{0.11}     \\ \hline
\texttt{einstein.en.txt}  & 467.63                   & 0.20      & 0.10      & \textbf{0.07}     \\ \hline
\texttt{english.001.2}    & 104.86                   & 4.17      & 2.41      & \textbf{0.55}     \\ \hline
\texttt{escherichiacoli}  & 112.69                   & 14.14     & 9.60     & \textbf{6.56}     \\ \hline
\texttt{fib41}            & 267.91                   & 0.03      & \textbf{0.00}      & 0.36     \\ \hline
\texttt{influenza}        & 154.81                   & 4.76      & 3.26      & \textbf{1.65}     \\ \hline
\texttt{kernel}           & 257.96                   & 2.37      & 1.10      & \textbf{0.82}     \\ \hline
\texttt{para}             & 429.27                   & 4.98      & 2.74      & \textbf{2.39}     \\ \hline
\texttt{proteins.001.1}   & 104.86                   & 4.13      & 2.64      & \textbf{0.59}     \\ \hline
\texttt{rs.13}            & 216.75                   & 0.02      & \textbf{0.00}      & 0.16     \\ \hline
\texttt{sources.001.2}    & 104.86                   & 4.10      & 2.34      & \textbf{0.45}     \\ \hline
\texttt{tm29}             & 268.44                   & 0.02      & \textbf{0.00}      & 0.72     \\ \hline
\texttt{world\_leaders}   & 46.97                    & 3.38      & 1.79      & \textbf{1.39}     \\ \hline
\end{tabular}
\end{table}

\begin{table}[tp]
\centering
\caption{Compression time regarding Pizzaz\&Chili repetitive corpus.}
\label{tab:compression-time}
\begin{tabular}{|l|c|c|c|c|c|}
\hline
\multicolumn{2}{|c|}{Pizza\&Chili Repetitive Corpus} & \multicolumn{3}{c|}{Compression Time (s)} \\ \hline
Experiment                & Input Size (MB)          & \our      & \repair   & \szip           \\ \hline
\texttt{cere}             & 461.29                   & \textbf{100.61}    & 464.62   & 693.10   \\ \hline
\texttt{coreutils}        & 205.28                   & \textbf{44.48}     & 210.21   & 85.19    \\ \hline
\texttt{dblp.xml.00001.1} & 104.86                   & \textbf{21.34}     & 71.85    & 25.63    \\ \hline
\texttt{dblp.xml.00001.2} & 104.86                   & \textbf{21.59}     & 72.31    & 25.60    \\ \hline
\texttt{dblp.xml.0001.1}  & 104.86                   & \textbf{21.21}     & 72.35    & 25.79    \\ \hline
\texttt{dblp.xml.0001.2}  & 104.86                   & \textbf{21.76}     & 73.70    & 27.16    \\ \hline
\texttt{dna.001.1}        & 104.86                   & \textbf{19.48}     & 73.83    & 63.56    \\ \hline
\texttt{einstein.de.txt}  & 92.76                    & 22.48		  & 62.17    & \textbf{16.26}    \\ \hline
\texttt{einstein.en.txt}  & 467.63                   & 135.19		  & 338.30   & \textbf{85.02}    \\ \hline
\texttt{english.001.2}    & 104.86                   & \textbf{27.79}     & 93.61    & 41.36    \\ \hline
\texttt{escherichiacoli}  & 112.69                   & \textbf{22.42}     & 138.06    & 143.05   \\ \hline
\texttt{fib41}            & 267.91                   & \textbf{15.58}     & 77.35    & 29.36    \\ \hline
\texttt{influenza}        & 154.81                   & \textbf{26.64}     & 108.98    & 46.14    \\ \hline
\texttt{kernel}           & 257.96                   & \textbf{60.26}     & 223.52   & 120.18   \\ \hline
\texttt{para}             & 429.27                   & \textbf{95.93}     & 512.93   & 583.92   \\ \hline
\texttt{proteins.001.1}   & 104.86                   & 29.05     	  & 82.86    & \textbf{21.27}    \\ \hline
\texttt{rs.13}            & 216.75                   & \textbf{12.04}     & 69.58    & 22.88    \\ \hline
\texttt{sources.001.2}    & 104.86                   & \textbf{23.56}     & 85.69    & 31.16    \\ \hline
\texttt{tm29}             & 268.44                   & \textbf{14.33}     & 92.70    & 39.11    \\ \hline
\texttt{world\_leaders}   & 46.97                    & \textbf{5.98}      & 23.57    & 9.26     \\ \hline
\end{tabular}
\end{table}

\begin{table}[tp]
\centering
\caption{Decompression time regarding Pizzaz\&Chili repetitive corpus.}
\label{tab:decompression-time}
\begin{tabular}{|l|c|c|c|c|}
\hline
\multicolumn{2}{|c|}{Pizza\&Chili Repetitive Corpus} & \multicolumn{3}{c|}{Decompression Time (s)} \\ \hline
Experiment                & Input Size (MB)          & \our      & \repair     & \szip       	  \\ \hline
\texttt{cere}             & 461.29                   & 18.88     & 13.31      & \textbf{2.01}   \\ \hline
\texttt{coreutils}        & 205.28                   & 13.53     & 3.95       & \textbf{2.37}    \\ \hline
\texttt{dblp.xml.00001.1} & 104.86                   & 5.61      & 0.82       & \textbf{0.34}    \\ \hline
\texttt{dblp.xml.00001.2} & 104.86                   & 5.62      & 0.85       & \textbf{0.34}    \\ \hline
\texttt{dblp.xml.0001.1}  & 104.86                   & 5.58      & 0.85       & \textbf{0.34}    \\ \hline
\texttt{dblp.xml.0001.2}  & 104.86                   & 5.65      & 1.04       & \textbf{0.34}    \\ \hline
\texttt{dna.001.1}        & 104.86                   & 6.31      & 1.75       & \textbf{0.37}    \\ \hline
\texttt{einstein.de.txt}  & 92.76                    & 5.57      & 0.45       & \textbf{0.29}    \\ \hline
\texttt{einstein.en.txt}  & 467.63                   & 29.40     & 2.70      & \textbf{1.43}    \\ \hline
\texttt{english.001.2}    & 104.86                   & 7.48      & 3.76       & \textbf{0.37}    \\ \hline
\texttt{escherichiacoli}  & 112.69                   & 7.49      & 3.36       & \textbf{0.87}    \\ \hline
\texttt{fib41}            & 267.91                   & 11.55     & \textbf{0.53}       & 1.09    \\ \hline
\texttt{influenza}        & 154.81                   & 9.16      & 1.09       & \textbf{0.67}    \\ \hline
\texttt{kernel}           & 257.96                   & 16.50     & 5.96       & \textbf{0.94}   \\ \hline
\texttt{para}             & 429.27                   & 19.18     & 12.88      & \textbf{2.16}    \\ \hline
\texttt{proteins.001.1}   & 104.86                   & 7.69      & 2.45      & \textbf{0.38}    \\ \hline
\texttt{rs.13}            & 216.75                   & 9.19      & \textbf{0.43}       & 0.71    \\ \hline
\texttt{sources.001.2}    & 104.86                   & 6.93      & 3.21       & \textbf{0.36}    \\ \hline
\texttt{tm29}             & 268.44                   & 10.26     & \textbf{0.53}       & 1.16    \\ \hline
\texttt{world\_leaders}   & 46.97                    & 1.66      & 0.45       & \textbf{0.20}    \\ \hline
\end{tabular}
\end{table}

\Section{Conclusions}\label{s:conclusion}

In the article we introduced a new grammar-based compression algorithm, called
\our, which is based on the induced suffix sorting framework of
\SAIS~\cite{Nong2009a}.
Experiments showed that \our is competitive compared to \repair and \szip, 
being very effective at handling repetitive strings.

\SubSection{Future works:}
As a future work, one can think of a \our/\repair hybrid approach
The key idea is to encode the first recursive levels
using \our and then shift to \repair. While making the compression a little
slower, this approach can make decompression faster while preserving a
good compression ratio.

We remark that \our, as well as \repair, can support extract random substrings
$T[l,r]$ without decompressing the complete string $T[1,n]$, by storing
additional data structures~\cite{Navarro2016}, whereas such operation is not
possible for LZ77 based compressors~\cite{Kreft2010}. We intend to implement
this operation aiming at reducing its memory footprint.
Also, an efficient way to search for a pattern in the compressed text is desirable.

\Section{References}
\bibliographystyle{IEEEtran}

\end{document}